      [1999/12/01 v1.4c Il Nuovo Cimento]
\documentclass{cimento}

\def\lesssim{\ \hbox{\raise 2pt \hbox{$<$} \kern -13pt
                     \lower 3pt \hbox{$\sim$}}\ }
\def\greatersim{\ \hbox{\raise 2pt \hbox{$>$} \kern -13pt
                     \lower 3pt \hbox{$\sim$}}\ }
\def\pythia{{\sc Pythia}}
\def\herwig{{\sc Herwig}}

\title{Production of jets at forward rapidities in hadronic collisions}
\author{
F.~Hautmann\from{ins:x}}
\instlist{\inst{ins:x}   Department of Theoretical Physics,  University of Oxford,   
Oxford OX1 3NP 
 }
\PACSes{
\PACSit{12.38,13.85,13.87}{\ldots}
}
\begin{document}

\maketitle

\begin{abstract}
We discuss  high-p$_{\rm{T}}$ production    processes 
at  forward   rapidities  in   hadron-hadron collisions,   and 
describe recent results  from  using QCD high-energy 
factorization in   forward  jet   production  at the LHC. 
\end{abstract}

\vskip -0.3 cm
\noindent {\em \hspace*{0.5 cm} Talk given at  Les  Rencontres de Physique de la 
Vall{\' e}e d'Aoste, La Thuile, 2009} 

\vskip -8.8 cm   
\hspace*{10.9 cm}  {\mbox{OUTP-09-20-P}}
\vskip 8.9 cm 

\section{Introduction} 

Experiments at the  Large Hadron Collider  (LHC) 
  will  explore  the  region of  large  rapidities  
 both with   general-purpose 
 detectors and with dedicated instrumentation, including 
  forward  calorimeters   and proton 
  taggers~\cite{cmsfwd,atlasfwd,fp420,cmstotem,aslano,grothe,heralhc}.     
The   LHC forward-physics   program 
involves   a wide range of topics, from  
new particle discovery  processes~\cite{fp420,fwdhiggs,fwdmssm}   
to  new  aspects of   
strong interaction physics~\cite{heralhc,denterria}  to 
heavy-ion  collisions~\cite{accardi03,heavy-ion-cmsnote}. 
Owing to  the  large center-of-mass energy    and 
  the unprecedented experimental coverage at large rapidities,  
 it becomes possible for the first time to investigate   the  forward region     
 with  high-$p_\perp$ probes.
 
In this article  we report on studies  of   forward  
production of jets~\cite{preprint}   based 
 on  QCD high-energy factorization   at 
fixed transverse momentum~\cite{hef}. This    theoretical 
framework serves  to  take into account 
consistently both the  higher-order  logarithmic corrections  
 in the large rapidity 
interval and those   in the hard jet transverse energy.    
In Sec.~2 we introduce  the basic structure of     
jet production in the LHC forward region.          
In Sec.~3 we consider   associated 
 parton showering effects.  In Sec.~4  we consider 
 effects from  the short-distance matrix 
 elements that control the resummation of logarithmically enhanced 
   corrections in $\sqrt{s} / E_T$, where $E_T$ is the hard jet transverse energy. 
 We give concluding remarks in Sec.~5.

\section{Forward jets at the LHC} 
 
\begin{figure}[htb]
\vspace{55mm}
\includegraphics{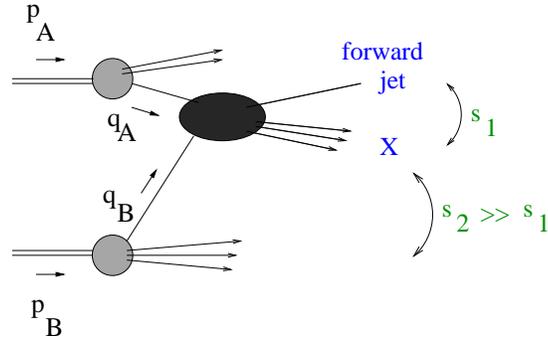}
\caption{Jet  production in the forward rapidity region 
in  hadron-hadron collisions.} 
\label{fig:forwpicture}
\end{figure}

\begin{figure}[htb]
\vspace{30mm}
\includegraphics{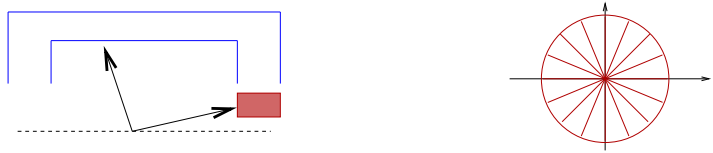}
\caption{(Left)  High-$p_\perp$  events  in the 
forward and central detectors; (right) azimuthal plane segmentation.} 
\label{fig:azimcorr}
\end{figure}

 The hadroproduction of a  forward jet   associated  
with  hard final state $X$  is pictured 
in Fig.~\ref{fig:forwpicture}.    
The kinematics 
  of the process     is  characterized 
by the  large  ratio  of sub-energies  $s_2  /  s_1 \gg 1 $   
 and  highly asymmetric longitudinal momenta in the partonic initial 
  state, $q_A \cdot p_B \gg q_B \cdot p_A$.  
 At the LHC the use of  forward calorimeters  allows  one to  
  measure    events  where   jet transverse momenta 
  $p_\perp  >   20$ GeV   are produced  several units of rapidity 
  apart,  $\Delta y   \greatersim 4 \div 6$~\cite{cmsfwd,aslano,heralhc}.  
 Working at    polar angles that are  small   but   sufficiently  far  from the beam axis 
 not to be affected by  beam remnants,     one measures 
 azimuthal plane correlations  
  between   high-$p_\perp$  events   (Fig.~\ref{fig:azimcorr})
  widely  separated    in rapidity~\cite{heralhc,preprint}.

The   presence of multiple   large-momentum scales  
implies   that, as      recognized in~\cite{muenav,vddtang,stirl94},    
reliable  theoretical predictions   for forward jets 
can only be obtained after  summing  
 logarithmic  QCD corrections at high energy
 to all orders in $\alpha_s$\footnote{Analogous observation applies to  
  forward jets  associated 
 to  deeply    inelastic  scattering~\cite{mueproc90c,forwdis92}.    Indeed, measurements of 
 forward jet cross sections at Hera~\cite{heraforw} have illustrated that 
 either fixed-order next-to-leading  
 calculations  or  standard shower 
 Monte Carlos~\cite{heraforw,web95,webetal99}, e.g.   \pythia\ 
  or \herwig,  are not  able to   describe 
 forward jet  $ep$ data.}.         
This    motivates    efforts~\cite{webetal99,orrsti,stirvdd,andsab}  to 
 construct   new,   improved  
  algorithms  for   Monte Carlo  event  generators capable of 
   describing   jet   production    beyond  the central rapidity region.

 In the  LHC forward   kinematics,  
  realistic  phenomenology  of      hadronic  jet final states 
  requires taking   account    of      
 both logarithms  of the large  rapidity  interval  
(of  high-energy type)  
and logarithms of the hard transverse momentum (of collinear type).  
The theoretical framework to resum   consistently 
both  kinds of logarithmic corrections  in QCD 
 calculations   is based on  high-energy    factorization at 
fixed transverse momentum~\cite{hef}.    

 Ref.~\cite{preprint}     
  investigates  forward jets  in this framework. It presents  the 
  short-distance matrix elements   needed to evaluate the 
   factorization formula, including  all partonic channels,  
   in a fully exclusive form.   
   On one hand, once convoluted with the BFKL off-shell gluon Green's function 
 according to the method 
 of~\cite{hef}, these matrix elements 
  control the summation of high-energy 
 logarithmic corrections to the jet cross sections. They 
 contain   contributions   both 
 to the   next-to-leading-order BFKL kernel~\cite{fadlip98} 
 and to the jet impact factors~\cite{mc98,schw0703}.  
   On  the other hand, they can   be used in a shower 
   Monte Carlo  generator    implementing  parton-branching kernels 
   at unintegrated level (see e.g.~\cite{jadach09,hj_ang}  for recent works)  
   to   generate fully exclusive events.

\begin{figure}[htb]
\vspace{45mm}
\includegraphics{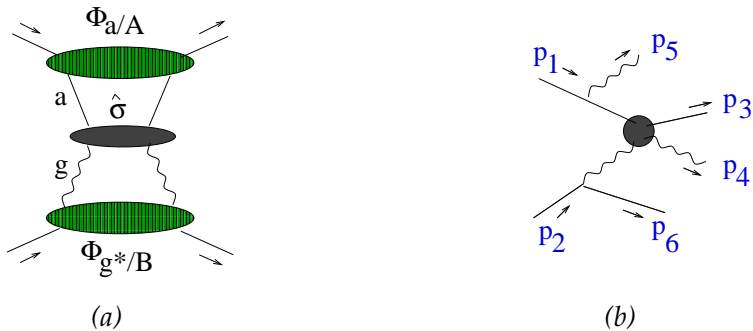}
\caption{(a) Factorized structure of the cross section; (b) a typical contribution  
 to the 
$q g$ channel matrix element.} 
\label{fig:sec2}
\end{figure}

The   high-energy  factorized 
form~\cite{preprint,hef,mc98}      
 of the  forward-jet  cross section is   represented   in Fig.~\ref{fig:sec2}a. 
Initial-state parton configurations  contributing to  
forward production are asymmetric, 
with the parton in the top subgraph being  probed near  the mass shell and  
large  $ x $,  
while  the parton in  the bottom subgraph is off-shell and small-$x$. 
The    jet  cross  section differential 
in the final-state   
transverse  momentum 
 $Q_\perp$  and  azimuthal angle $\varphi$ 
is given  schematically  by~\cite{preprint,hef,mc98}  
\begin{equation}
\label{forwsigma}
   {{d   \sigma  } \over 
{ d Q_\perp^2 d \varphi}} =  \sum_a  \int  \    \phi_{a/A}  \  \otimes \  
 {{d   {\widehat  \sigma}   } \over 
{ d Q_\perp^2 d \varphi  }}    \  \otimes \   
\phi_{g^*/B}    \;\; , 
\end{equation}
where  
$\otimes$ specifies  a convolution in both longitudinal and transverse momenta, 
$ {\widehat  \sigma} $  is the  hard scattering  cross section,  calculable 
 from  a  suitable off-shell continuation of 
perturbative   matrix elements,  $ \phi_{a/A} $ is the distribution of parton 
$a$ in hadron $A$ 
obtained from   near-collinear shower evolution, and $ \phi_{g^*/B} $ is  
  the gluon unintegrated distribution in hadron $B$ 
  obtained from non-collinear, 
  transverse momentum  dependent shower evolution. 
 
In the  next section we  comment on the initial-state shower evolution. 
In Sec.~4 we turn to   hard-scattering  contributions.

\section{Parton shower evolution} 

\begin{figure}[htb]
\vspace{45mm}
\includegraphics{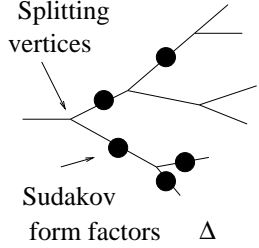}
\caption{Parton  branching in terms of splitting probabilities and form factors.}
\label{fig:pshower}
\end{figure}

Parton distributions can be obtained  
by parton-shower Monte Carlo methods   
  via   branching algorithms 
based  on collinear evolution 
of the jets developing    
from  the hard event~\cite{mc_lectures}. The branching probability  can be   given 
 in terms    of  two basic 
quantities  (Fig.~\ref{fig:pshower}), the  splitting functions   
 at the vertices of the parton cascade     and the  form 
 factors   to go from one vertex  to the other. 
An important ingredient of this approach is the inclusion of  soft-gluon 
coherence  effects~\cite{mc_lectures,dokrev,mc89} through angular ordering of the 
emissions in the shower.   

\begin{figure}[htb]
\vspace{36mm}
\includegraphics{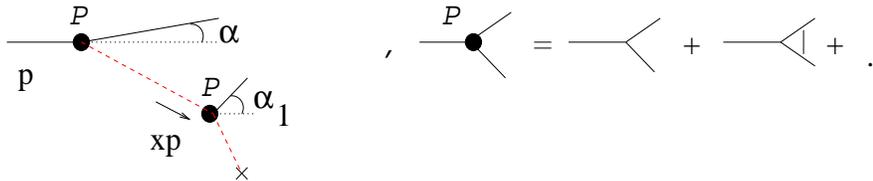}
\caption{(left) Coherent radiation 
 in the space-like parton shower for $x \ll 1$; (right) the unintegrated 
splitting function ${\cal P}$, including small-$x$ virtual 
corrections.} 
\label{fig:coh}
\end{figure}

Corrections to collinear-ordered showers, however,  arise 
 in  high-energy processes with multiple  
 hard scales~\cite{heralhc,mw92,bo04},  as is the case  with 
 the  production  of jets at  forward rapidities in Fig.~\ref{fig:forwpicture}.
 In particular, new  color-coherence   effects  set in  in this regime 
 due to    emissions  from    internal lines in the  
 branching decay chain~\cite{heralhc,mc98,anderss96}  
    that  involve
 space-like partons   
 carrying  small   longitudinal momentum   
 fractions.   
 The picture of the coherent branching   is modified   in this case because 
 the emission  currents become dependent on the total transverse 
 momentum  transmitted down the  initial-state parton decay 
   chain~\cite{hef,mc98,mw92,bo04,jung04}.  Correspondingly,   
   one needs to work 
   at the level of    unintegrated   
   splitting functions and  partonic distributions~\cite{jcc-lc08,hj_rec}  
   in order to take into account 
     color coherence not only  for  large 
 $x $ but also  for  small $x$  in the angular region (Fig.~\ref{fig:coh}) 
\begin{equation}
\label{cohregion}
\alpha / x  > \alpha_1 > \alpha \hspace*{0.3 cm}    , 
\end{equation} 
where the angles  $\alpha$ for the 
partons radiated from the initial-state 
shower are taken with respect to the 
initial beam jet direction, and increase with increasing off-shellness.

The case of  LHC forward  jet production 
 is a multiple-scale problem where 
coherence effects  of the kind  above  
enter, in the factorization formula  (\ref{forwsigma}), 
 both the  short-distance factor $ {\widehat  \sigma} $ and   the 
long-distance factor  $\phi$.
 Contributions   from  the coherence   region   
(\ref{cohregion})  are potentially enhanced by   terms 
$ 
\alpha_s^n \ln^{m}  \sqrt{s} / p_\perp   $ 
where $\sqrt{s}$ is the total center-of-mass energy and $p_\perp$  is  
the   jet  transverse momentum\footnote{Terms with 
 $m  > n$   are known to 
   drop out  from inclusive  processes due to strong cancellations 
associated with coherence,    
   so that,  for instance,  the 
    anomalous dimensions $\gamma^{ i j}$  for   
 space-like evolution    receive   at most   single-logarithmic 
 corrections at high energy~\cite{fadlip98,ch94}. This need not be the case for  
 exclusive  jet  distributions, where   such cancellations  are not  present 
and one may expect   larger enhancements.}.  
These contributions represent corrections to 
the  angular  ordering  implemented in  collinear  showers and 
are not included   at present 
in standard Monte Carlo generators~\cite{mc_lectures}. 
 Work  to develop 
 methods for   unintegrated shower evolution, capable of including 
 such corrections,  is 
   underway  by  several authors. 

The  proposal~\cite{jadach09} incorporates 
NLO  corrections to  flavor non-singlet  QCD evolution  in 
  an unintegrated-level Monte Carlo. 
The approach is based on the generalized ladder expansion of~\cite{CFP}, which is 
   extended to the high-energy   
region  in~\cite{ch94}. This approach   could  in principle   be applied generally,  
including  flavor  singlet evolution, and used 
to treat also forward hard processes.

Shower Monte Carlo generators 
 based on small-x evolution equations,  
 on  the other hand,   have  typically included  the unintegrated 
 gluon distribution only~\cite{bo04,jung04}.    We observe that 
  unintegrated quark contributions  can be incorporated    
   for sea quarks    via  the 
  transverse-momentum dependent   but universal
 splitting kernel given in~\cite{ch94}, which has the structure    
 \begin{equation} 
 \label{pqg}
{\cal P}_{g \to q}  ( z;  q_\perp,  k_\perp )  =    P_{qg}^{(0)}  ( z) \ 
\left( 1 +  \sum_{n=1}^{\infty}   \ b_n (z )   ( k_\perp^2 /  q_\perp^2 )^n  \right)  \;\;\; , 
\end{equation}  
where $P^{(0)} $  is  the DGLAP  splitting function, and all coefficients 
$b_n$ are known. 
The kernel (\ref{pqg}) has been  used for inclusive small-x calculations~\cite{cc06}. 
Its  Monte Carlo  implementation is relevant to take  
into account  effects from  the   unintegrated quark  distribution 
  in   the simulation  of   exclusive final states~\cite{unint-quark}. 
We note that 
quark contributions to  evolution  
 at the fully unintegrated level 
  will   also enter  the  treatment of  the 
    subleading high-energy corrections   that are 
    being   discussed for  jet production~\cite{schw0703,mn_pheno}. 

    Analyses of  forward jet hadroproduction     including parton showering  effects 
     are in progress~\cite{prepar}.

\section{The factorizing hard cross sections}

 Logarithmic corrections 
 for large rapidity $y \sim  \ln s / p_\perp^2$ 
are  resummed to all orders in $\alpha_s$ 
 via Eq.~(\ref{forwsigma}),    
   by convoluting     (Fig.~\ref{fig:sec2})  unintegrated distribution 
 functions  with   
 well-prescribed short-distance matrix elements, 
 obtained from the high-energy limit of  higher-order 
 scattering amplitudes~\cite{preprint,mc98}.  
With reference to     Fig.~\ref{fig:sec2}b, 
in the   forward production region we have 
  $(p_4+ p_6)^2  \gg  (p_3 +p_4)^2   $     and 
 longitudinal momentum ordering,  so that 
\begin{equation}
\label{fwdkin}
p_5 \simeq  (1 - \xi_1 ) p_1 \;\;\;   ,  \;\;\;\;\;  p_6 \simeq   (1 - \xi_2 ) p_2 - k_\perp   
 \;\;\;   ,  \;\;\;\;\;  
\xi_1 \gg \xi_2     \;\;  .  
\end{equation}
Here $\xi_1$ and $\xi_2$ are longitudinal momentum fractions,   and $ k_\perp $  is the 
di-jet  transverse momentum in the laboratory frame.   
It is convenient to  define  the rapidity-weighted average
$Q_\perp = (1-\nu) p_{\perp  4} - \nu p_{\perp  3}$, with  
$\nu =  (p_2 \cdot p_4) / p_2 \cdot (p_1 -p_5) $.  In    Fig.~\ref{fig:sec2}b  
Eq.~(\ref{forwsigma})  factorizes  the high-energy $q g$ amplitude 
 in front       of the (unintegrated) distribution from 
 the splitting in the bottom subgraph. 
The factorization in terms of   this    parton 
splitting    distribution  is valid at large $y$      not only in  the collinear  region 
but also in the large-angle emission region~\cite{hef}.    As a result  
  the rapidity resummation is  carried out  consistently with 
perturbative  high-$Q_\perp$ corrections~\cite{hef,mc98} at any fixed 
order in $\alpha_s$.

\begin{figure}[htb]
\vspace{44mm}
\includegraphics{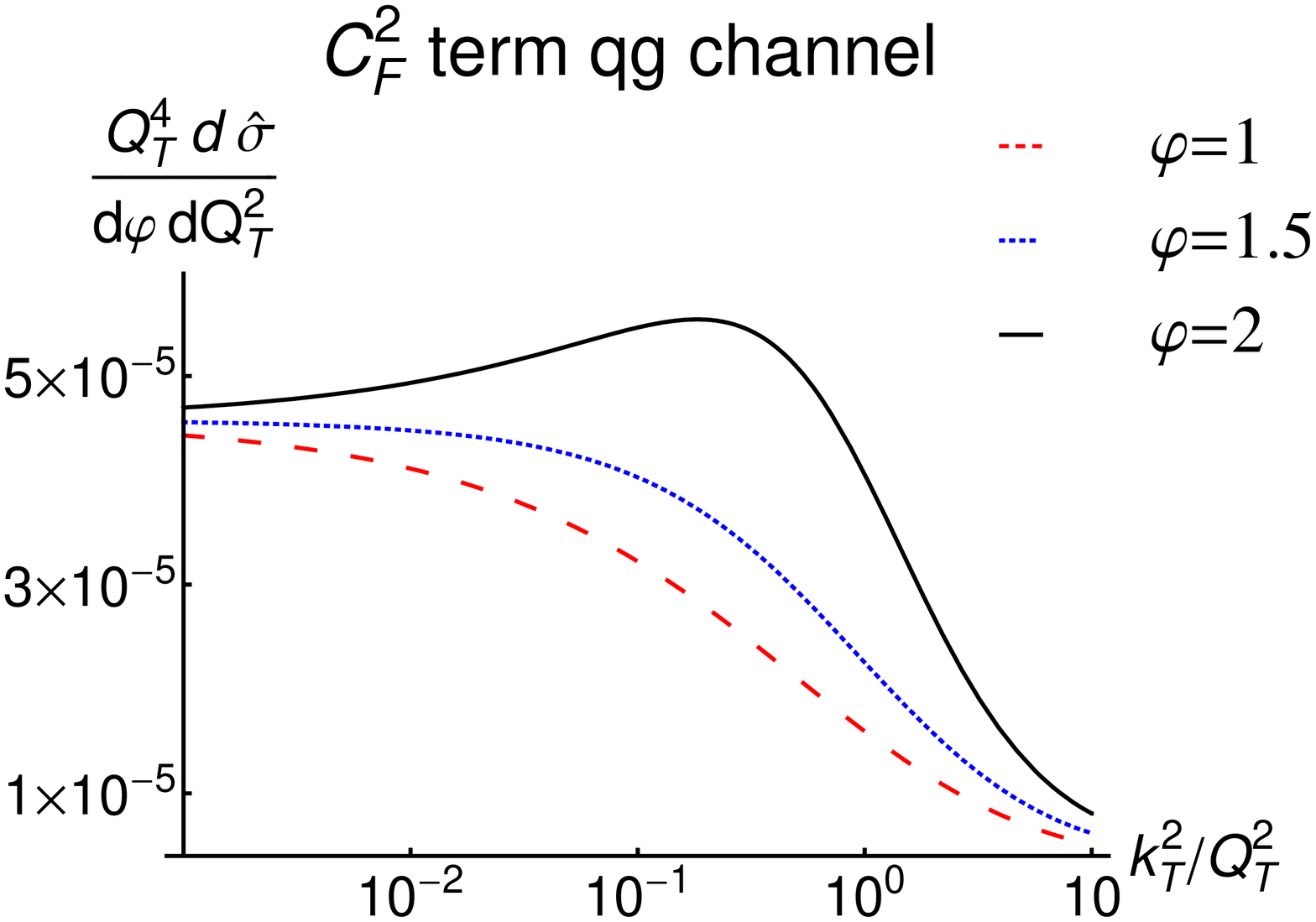}
\includegraphics{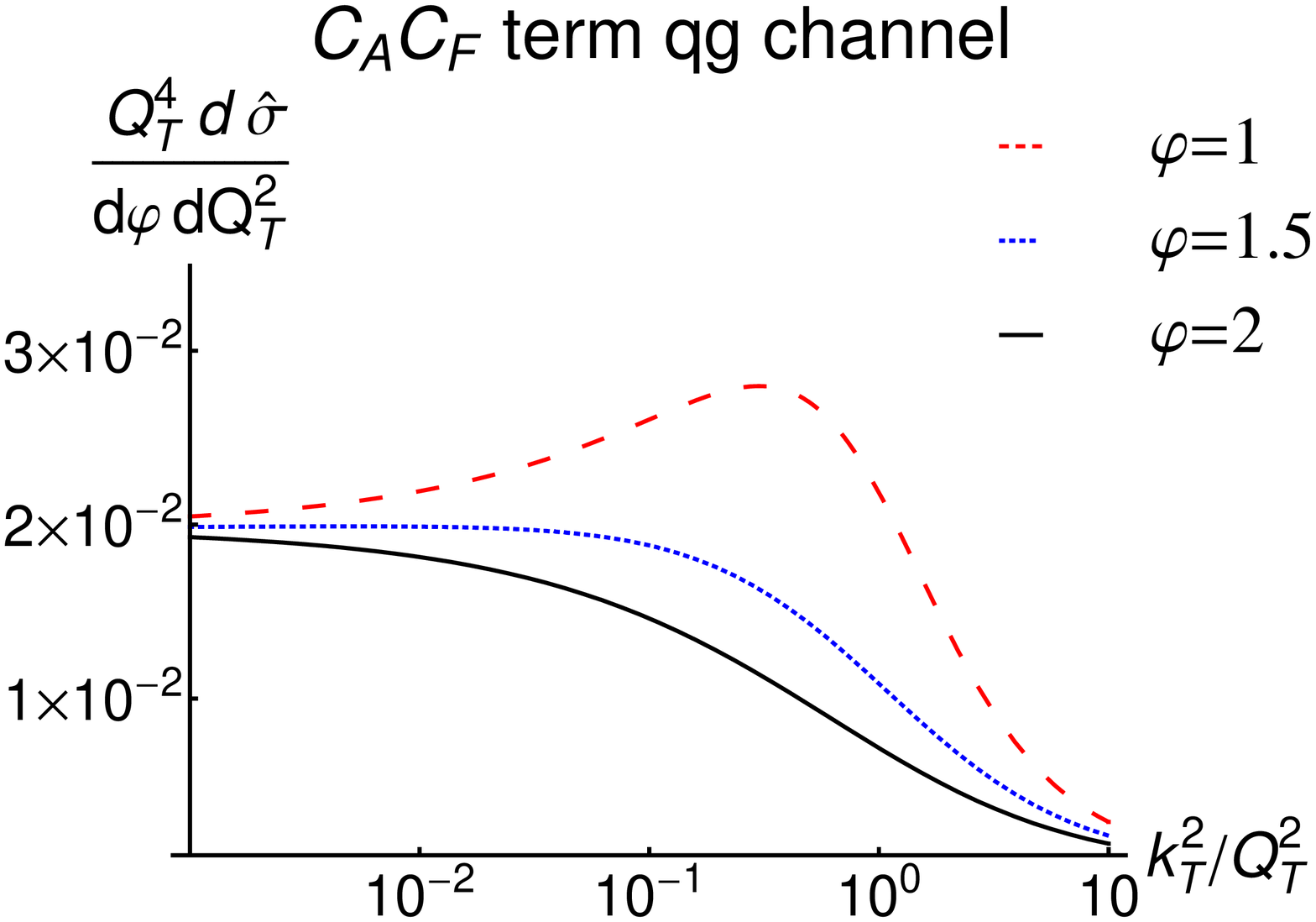}
\caption{The  $k_T / Q_T$  dependence of the  
factorizing   $q g$   hard  cross 
section at high energy~\cite{preprint}:           
 (left) $C_F^2$ term;  (right) $C_F   
C_A$ term.} 
\label{fig:forwplot}
\end{figure}

\begin{figure}[htb]
\vspace{44mm}
\includegraphics{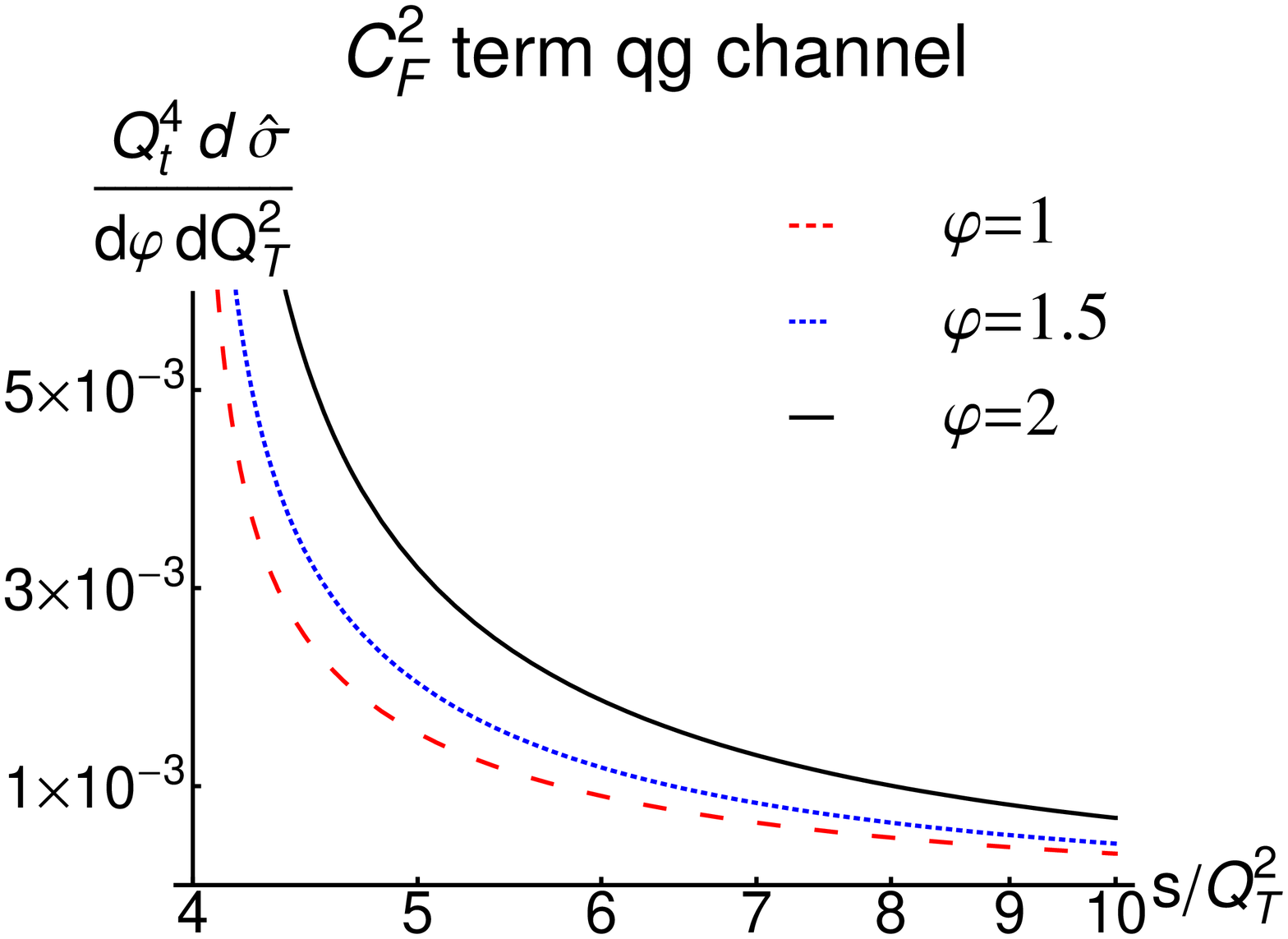}
\includegraphics{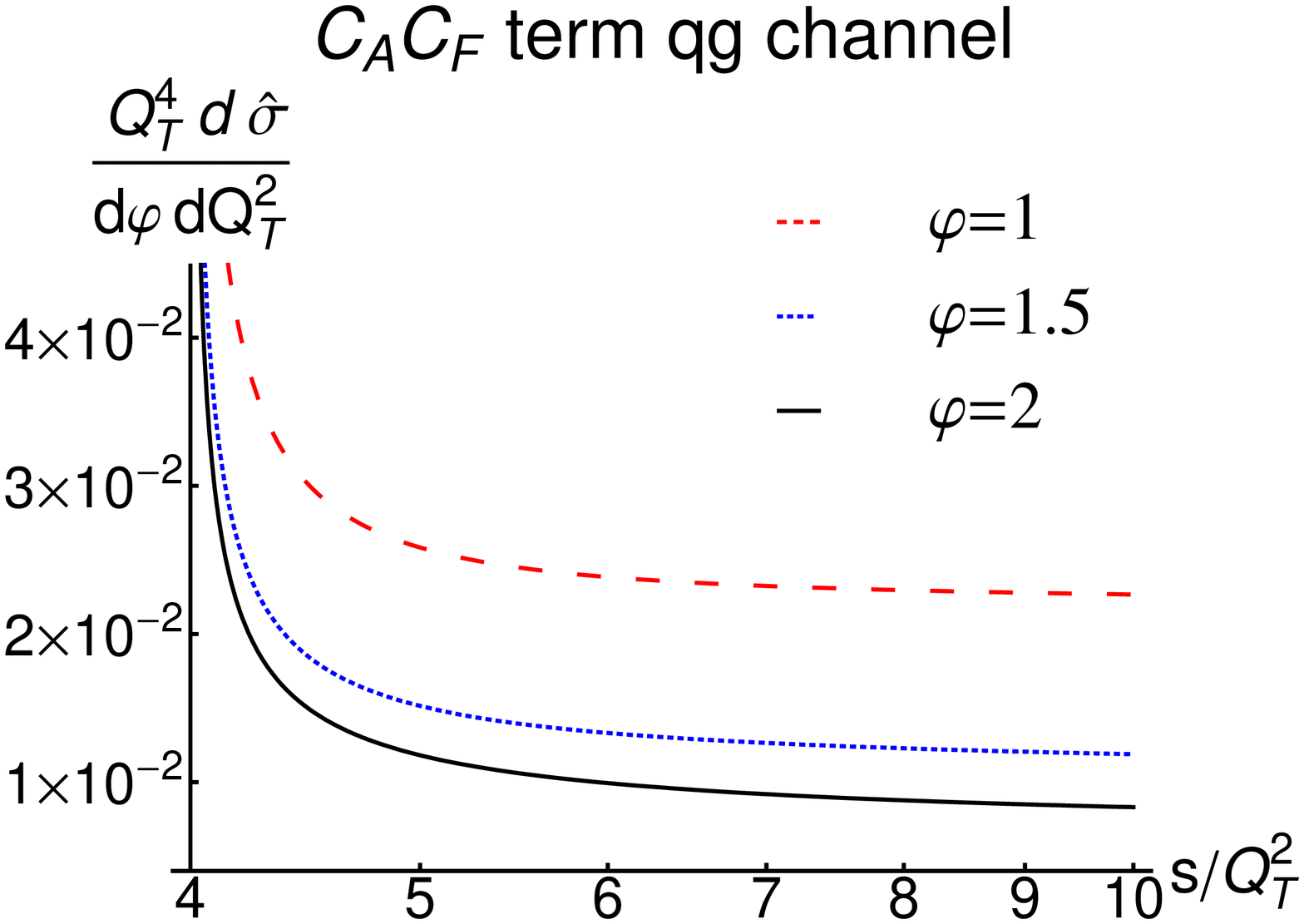}
\caption{The  energy dependence of the    $q g$ hard cross section~\cite{preprint} 
($  k_T / Q_T =1$).} 
\label{fig:forwplot1}
\end{figure}

The explicit expressions  for  the relevant 
high-energy  amplitudes  
are  given in~\cite{preprint}. 
Figs.~\ref{fig:forwplot}  and \ref{fig:forwplot1}  illustrate  features of the 
factorizing matrix elements, partially integrated over final states. We  plot  
distributions differential in  $Q_\perp$ and  azimuthal angle $\varphi$ 
($\cos \varphi = Q_\perp \cdot k_\perp / |  Q_\perp |  | k_\perp | $) for the case 
of the $q g$ channel.   
Fig.~\ref{fig:forwplot} shows the dependence on  $k_\perp$,  which measures 
  the distribution  
  of   the  third    jet   recoiling against the leading di-jet system. 
    Fig.~\ref{fig:forwplot1}   shows the energy dependence.

The region    $ k_\perp / Q_\perp   \to  0$  in Fig.~\ref{fig:forwplot}  
corresponds to 
the leading-order   process with two back-to-back 
jets.   
The resummation  of  the higher-order  logarithmic corrections 
 for large  $y \sim  \ln s / p_\perp^2$  is precisely  
 determined~\cite{hef,mc98}    by  integrating the 
 u-pdfs   over the  $ k_\perp$-distribution  in  Fig.~\ref{fig:forwplot}.   
So the results       in   
Fig.~\ref{fig:forwplot}  
 illustrate    quantitatively  the  significance of  
contributions  
with   $k_\perp \simeq  Q_\perp$ in the large-$y$   region.    
  The role of coherence  from multi-gluon emission 
  is to set the  dynamical cut-off     at values of 
  $ k_\perp $ of order  $ Q_\perp $. 
  Non-negligible 
  effects arise at high energy    from the finite-$k_\perp $  tail. 
These  effects are not included in  collinear-branching  
 generators  (and only partially in fixed-order 
 perturbative calculations),   
 and  become  more and more important  as  the jets are  observed at 
large rapidity separations.   The   dependence on the azimuthal angle 
in    Figs.~\ref{fig:forwplot}  and \ref{fig:forwplot1} 
is  also relevant, as   forward  jet    measurements  will rely 
  on azimuthal plane correlations between  jets  
far apart  in rapidity (Fig.\ref{fig:azimcorr}). 
  
  Results for  all  other partonic channels   are given    in~\cite{preprint}.   
After including parton showering~\cite{prepar},    quark- 
  and gluon-initiated  contributions   are of comparable size 
 in  the 
  LHC forward kinematics:  
    realistic phenomenology requires including all channels. 
  Note  also that 
since the forward kinematics selects 
asymmetric parton momentum fractions,   effects   
     due to    the   $ x \to 1$   endpoint   behavior~\cite{fhfeb07}   
  at  the 
fully  unintegrated level  may become relevant as well.  

 Let us finally  recall that    if  effects of    high-density parton 
dynamics~\cite{denterria,ianmue}   show up 
 at the LHC,  they will   influence 
 forward jet event distributions.  In such a case,   
 the unintegrated  formalism  discussed above 
 would likely be  the   natural  framework  
 to implement   this  dynamics  at   parton-shower level.

\section{Conclusion}

Forward + central detectors  at the LHC  
 allow  
jet   correlations   to be  measured 
across  rapidity intervals of  several  units,   
 $\Delta y   \greatersim 4 \div 6$.      
   Such multi-jet    states   can   be relevant to 
   new particle  discovery processes as well as 
    new aspects   of  standard model physics.

Existing sets of  forward-jet  data in ep collisions, much more limited  
    than the potential LHC  yield, 
  indicate  that neither conventional parton-showering Monte Carlos 
nor next-to-leading-order   QCD calculations  are capable of  describing 
  forward jet    phenomenology.  
 Improved methods to evaluate QCD    predictions 
 are needed to treat the multi-scale 
 region implied by  the  forward kinematics.   

In this article 
we have discussed ongoing progress,      
examining in particular   factorization properties 
of multi-parton matrix elements in the forward region, 
and prospects  
to include  parton-showering effects 
with gluon coherence     not only in the collinear region 
but also   in the large-angle emission region.

\acknowledgments  
 I  thank    M.~Greco,  the  conference organizers   and    the conference staff   
 for  the kind    invitation  and  for   the  
  nice   atmosphere at the meeting. 
The results presented in this article have been obtained   
in collaboration with 
M.~Deak, H.~Jung and K.~Kutak.

\end{document}